\documentclass[final,3p,times,sort&compress]{elsarticle}
\usepackage{lineno,hyperref}
\usepackage{graphicx}
\usepackage{dcolumn}
\usepackage{bm}
\usepackage{upgreek}
\usepackage{hyperref}
\usepackage{ams math}
\usepackage{multirow}

\usepackage{multicol} 

\usepackage{nomencl} 

\makenomenclature

\let\oldequation\equation
\let\oldendequation\endequation
\renewenvironment{equation}
  {\linenomathNonumbers\oldequation}
  {\oldendequation\endlinenomath}

\usepackage{etoolbox}
\renewcommand\nomgroup[1]{%
  \item[\bfseries
  \ifstrequal{#1}{G}{Greek symbols}{%
  \ifstrequal{#1}{S}{Subscripts}{}}%
]}

\newcommand{\Nu}{Nu}%
\newcommand{\Ca}{Ca}%
\newcommand{\We}{W\!e}%
\newcommand{\Ja}{Ja}%
\journal{Applied Thermal Engineering}

\begin{document}

\begin{frontmatter}

\title{Growth of Elongated Vapor Bubbles During Flow Boiling Heat Transfer in Wavy Microchannels}

\author{Odumuyiwa A. Odumosu}
\author{Huashi Xu}
\author{Tianyou Wang}
\author{Zhizhao Che\corref{cor1}}
\cortext[cor1]{Corresponding author.
}
\ead{chezhizhao@tju.edu.cn}
\address{State Key Laboratory of Engines, Tianjin University, Tianjin, 300350, China.}


%
%

\begin{abstract}
Flow boiling within microchannels is an efficient heat removal mechanism by utilizing the latent heat of the fluid. Numerical simulations can resolve the small temporal and spatial scales of the vapor bubbles in the flow boiling process, thus accessing many details of the flow which is unattainable experimentally. Most of the previous studies of flow boiling heat transfer used straight microchannels, even though curved microchannels have potential to further enhance the heat transfer performance. In this study, the flow boiling heat transfer in wavy microchannels is studied via numerical simulation by focusing on the growth of the vapor bubbles. Straight and wavy microchannels are considered for comparison and analysis. The numerical simulations are conducted using the finite volume method to solve the flow and the heat transfer, and the volume of fluid (VOF) method to capture the interface dynamics. The effects of key dimensionless numbers on bubble growth rate and wall heat transfer are quantified, including the Weber number, the Capillary number, and the Jakob number. The results indicate that the wavy structure of the microchannel has a strong effect on the bubble growth and the wall heat transfer in microchannels. As the waviness of the microchannel increases, the bubble grows faster because of the bubble deformation by the wavy channel, the bubble moves faster due to the bubble expansion, and the heat transfer is enhanced by the perturbation of the bubble to the flow. With the large velocity perpendicular to the channel flow direction induced by the waviness, the local Nusselt number of the wavy channel could be high up to 2.6 times higher than the straight channel.
\end{abstract}

\begin{keyword}
\texttt {Flow boiling \sep Wavy microchannel \sep Phase change \sep Heat transfer enhancement}
\end{keyword}

\end{frontmatter}


\def \scaleSize {0.8}
\def \scaleSiz2 {0.6}

\section{Introduction}\label{sec:sec01}
The application of micro-structured devices in cutting-edge technologies has led to the demand for the removal of high heat flux. One of the prominent breakthroughs in this field is the introduction of narrow channels. A narrow passage or microchannel provides a large heat-transfer area per unit volume of fluid flow, which helps in effective heat removal. Further, phase-change flow enhances the heat-removal capacity by utilizing the latent heat of the fluid \cite{Jain2020BubbleGrowth} and the flow vortices induced by the presence of the interface \cite{Che2015HeatTransfer3Ddroplet}. Due to the excellent heat dissipation capability, microchannel heat sinks designed for micro-structured devices have been widely studied especially in microelectronic devices \cite{Bertsch2008ReviewFlowBoiling, Garimella2006ThermalManagement}. Pertaining to this, numerous experimental and numerical studies have been carried out to understand flow boiling in microchannels focusing on flow boiling heat transfer mechanism \cite{Fayyadh2017FlowBoiling, Magnini2017SlugFlowBoiling, Magnini2016FlowBoiling}, bubble dynamics, and flow pattern transition \cite{Chen2020ExperimentalFlowBoiling, Liu2017BubbleTrainMicrochannel, Prajapati2017FlowBoiling}.

Flow boiling in a microchannel can be classified into various stages according to the flow patterns such as single-phase, bubbly, confined bubble (slug/plug), annular, and dry-out regimes based on the heat flux and mass flux as observed by Harirchian and Garimella \cite{Harirchian2012MicrochannelFlowBoiling}. Within a microchannel, the entire cross-section of the channel is rapidly filled by a growing bubble. During bubbly flow, the discrete bubbles coalesce and produce large bubbles, while the slug and annular flow regimes occupy a larger area of the flow map \cite{Revellin2006DiabaticTwoPhaseFlow}. In particular, the slug (elongated bubble) flow regime has been regarded as an optimal operating condition for microchannel heat exchangers, due to its efficient heat transfer \cite{Ferrari2018SlugFlowBoiling}. Many researchers have studied experimentally the slug flow regime during flow boiling in single channels \cite{Bigham2015FlowBoiling, Gedupudi2011BubbleGrowth, Huh2007FlowBoiling, Wang2014MicrochannelFlowBoiling}. The small cross-sectional dimensions of the channel results in the confinement of the vapor bubble and its growth in the axial direction forming an elongated bubble with the thin film surrounding the vapor bubble \cite{Jain2020BubbleGrowth}.

Apart from experiments, numerical simulation is an important tool to access the details of flow and heat transfer of flow boiling in microchannels. Yang et al.\ \cite{Yang2008TwoPhaseFlowBoiling} simulated the flow boiling of R141b in a horizontal coiled tube with the empirical phase change rate parameter model of Lee \cite{Lee1980bookChapter} based on the volume of fluid (VOF) method, and the corresponding experimental heat transfer measurements, as well as flow pattern visualization, were also conducted and compared \cite{Yang2008TwoPhaseFlowBoiling}. Mukherjee and Kandlikar \cite{Mukherjee2005BubbleGrowth} numerically studied the growth of vapor bubbles during flow boiling of water in a square microchannel with superheated fluid and constant wall temperature using the level set (LS) method. Mukherjee et al.\ \cite{Mukherjee2011BubbleGrowth} further studied the effects of wall superheat, liquid mass flux, surface tension, and contact angle on bubble growth and heat transfer characteristics. Zhuan and Wang \cite{Zhuan2012FlowBoilingMicrochannel} studied numerically the flow pattern transitions for saturated boiling of R134a and R22 in circular microchannels, and investigated the bubble growth and coalescence of bubbly flow, slug flow, and annular flow. Luo et al.\ \cite{Luo2017BubbleGrowth} combined the VOF method and the saturation temperature recovery model \cite{Rattner2014VolumeOfFluidFormulation} to study the effects of boundary conditions such as the heat flux, contact angle, surface tension, and inlet Reynolds number on the bubble dynamics for the growth of single nucleated bubble and coalescence of adjacent bubbles. Ferrari et al.\ \cite{Ferrari2018SlugFlowBoiling} performed a numerical investigation on flow boiling heat transfer of R245fa in a square microchannel in the slug flow regime, and the effects of the channel's cross-sectional geometry (circular or square) were studied.

Even though many experimental studies and numerical simulations have been devoted to flow boiling heat transfer in microchannels, most of them used straight microchannels. In contrast, curvatures in microchannels can induce dean flow \cite{Dean1928CurvedChannel} and vortex asymmetry \cite{Che2010}, which is potential to further enhance the heat transfer performance. Therefore, the objective of the present work is to study the flow boiling heat transfer in wavy microchannels. We will focus on the growth of the elongated vapor bubbles and its effect on heat transfer performance. Further, the effects of key dimensionless parameters on bubble growth and corresponding wall heat transfer are to be analyzed and explained, including the Weber number, the Capillary number, and the Jakob number.

\section{Numerical method}\label{sec:sec02}
In the present study, the open-source platform OpenFOAM is used to perform the flow boiling simulation. The interThermalPhaseChangeFoam solver by Nabil and Rattner \cite{Nabil2016InterThermalPhaseChangeFoam} is adopted for the simulation. The thermal phase-change solver is extended from the adiabatic two-phase flow solver interFoam, which is described in detail in Ref.\ \cite{Deshpande2012TwoPhaseFlowSolver}. The code solves conservation equations for mass (via a pressure Poisson equation), momentum, thermal energy, and the phase fraction.

\subsection{Governing equations}\label{sec:sec021}
In this section, the details of mathematical models and numerical methods for this study are presented. The VOF method is employed to capture the interface dynamics. The single-fluid set of equations for immiscible and incompressible phases under flow boiling conditions can be summarized as follows. The mass conservation equation is
\begin{equation}\label{eq:eq01}
  \nabla \cdot \mathbf{u}={{\dot{v}}_{lv}}
\end{equation}
where $\mathbf{u}$ is the velocity vectors for both phases, ${{\dot{v}}_{lv}}$ the dilatation rate for phase change (see Eq.\ (\ref{eq:eq12}) in the phase-change model). The momentum conservation equation is
\begin{equation}\label{eq:eq02}
  \frac{\partial (\rho \mathbf{u}\text{)}}{\partial t}+\nabla \cdot (\rho \mathbf{uu}\text{)}=-\nabla p+\nabla \cdot \{\mu [\nabla \mathbf{u}+{{\text{(}\nabla \mathbf{u}\text{)}}^{T}}]\}+{{\mathbf{f}}_{\sigma }}
\end{equation}
where $p$ is the pressure, $\rho$ and $\mu$ are the density and the viscosity of the fluid, and ${{\mathbf{f}}_{\sigma }}$ is the surface tension force (see Eq.\ (\ref{eq:eq08}) in the continuum surface force (CSF) model). The energy conservation equation is
\begin{equation}\label{eq:eq03}
\frac{\partial (\rho i)}{\partial t}+\nabla \cdot (\rho \mathbf{u}i)=\nabla \cdot (k\nabla T)-{{\dot{q}}_{lv}}
\end{equation}
where $i$ is the enthalpy, $k$ is the thermal conductivity, and ${{\dot{q}}_{lv}}$ is the phase-change heat source term (see Eq.\ (\ref{eq:eq11}) in the phase-change model). The mass-average enthalpy $i$ can be written as
\begin{equation}\label{eq:eq04}
i=\frac{(1-\alpha ){{\rho }_{v}}{{c}_{p,v}}+\alpha {{\rho }_{l}}{{c}_{p,l}}}{\rho }(T-{{T}_{ref}})
\end{equation}
where ${{\rho }_{v}}$ and ${{\rho }_{l}}$ are the density of the vapor and the liquid, ${{c}_{p,v}}$ and ${{c}_{p,l}}$ are the constant pressure specific heat capacity of the vapor and the liquid, $T$ is the temperature of the fluid and ${{T}_{ref}}$ is reference temperature. The VOF equation is
\begin{equation}\label{eq:eq05}
\frac{\partial \alpha }{\partial t}+\nabla \cdot (\mathbf{u}\alpha \text{)}={{\dot{\alpha }}_{lv}}
\end{equation}
where $\alpha$ is the liquid phase-fraction field, and ${{\dot{\alpha }}_{lv}}$ is phase fraction generation (see Eq.\ (\ref{eq:eq13}) in the phase-change model). The mesh cell is full of liquid when $\alpha = 1$, it is full of vapor when $\alpha = 0$, and $0 < \alpha <1$ indicates a mixture of liquid and vapor.

Since the working fluid used in this study is considered incompressible, constant fluid thermophysical properties are adopted, including the thermal conductivity, the viscosity, and the density. The density is evaluated using the arithmetic mean of the liquid and vapor phases
\begin{equation}\label{eq:eq06}
\rho =(1-\alpha ){{\rho }_{v}}+\alpha {{\rho }_{l}}.
\end{equation}
For the dynamic viscosity and the thermal conductivity near the interface, a weighted average of the arithmetic mean and the harmonic mean is used based on the orientation of the interface (see Ref.~\cite{Nabil2016InterThermalPhaseChangeFoam} for the details). This method can improve the accuracy of the viscous stress and the thermal resistance near the interface, which is important in determining the flow and heat transfer near the interface.


Surface tension force is important in the flow boiling process and is considered in the momentum equation. The surface tension force on liquid--vapor interface is regarded as an additional source term ${{\mathbf{f}}_{\sigma }}$ in Eq.\ (\ref{eq:eq02}) according to Brackbill's continuum surface force (CSF) model \cite{Brackbill1992ModelingSurfaceTension}. A multiplier term $2\rho /\left( {{\rho }_{v}}+{{\rho }_{l}} \right)$ is employed so that the surface tension source term is proportional to the average density \cite{Brackbill1992ModelingSurfaceTension}.
\begin{equation}\label{eq:eq08}
{{\mathbf{f}}_{\sigma }}=\sigma \kappa \mathbf{n}\left| \nabla \alpha  \right|\frac{2\rho }{{{\rho }_{v}}+{{\rho }_{l}}}
\end{equation}
where $\kappa $ and $\mathbf{n}$ are the interface curvature and the unit vector normal to the interface.
\begin{equation}\label{eq:eq09}
\kappa =-\nabla \cdot \mathbf{n}
\end{equation}
\begin{equation}\label{eq:eq10}
\mathbf{n}=\frac{\nabla \alpha }{\left| \nabla \alpha  \right|}
\end{equation}

\subsection{Phase change model}\label{sec:sec022}
The adopted phase-change model is based on that of Rattner and Garimella \cite{Rattner2014VolumeOfFluidFormulation}. This formulation operates on mesh connectivity (which cells share faces) and volumetric field data, where geometric interface reconstruction is not required as in Refs.\ \cite{Yang2008TwoPhaseFlowBoiling, Yuan2008VolumeOfFluidMethod}. The phase-change source term (${{\dot{q}}_{lv}}$ in Eq.\ (\ref{eq:eq03})) is scaled such that after each time step ($\Delta t$), the cells containing the liquid-vapor interface return to the equilibrium temperature ($T_{sat}$). This model has been verified to be accurate for overall heat transfer predictions, provided that the mesh resolution in the vicinity of the interface is sufficient \cite{Onishi2013HeatTransferMinichannel, Rattner2014VolumeOfFluidFormulation} and neglects in-cell thermal resistance. The interface-cell contributions to overall thermal resistance are generally negligible in a two-phase flow simulation.
\begin{equation}\label{eq:eq11}
{{\dot{q}}_{lv}}=\frac{\rho {{c}_{p}}(T-{{T}_{sat}})}{\Delta t}
\end{equation}
Once ${{\dot{q}}_{lv}}$ is obtained for interface cells, the dilatation (${{\dot{v}}_{lv}}$) and phase fraction generation (${{\dot{\alpha }}_{lv}}$) source terms are evaluated as
\begin{equation}\label{eq:eq12}
{{\dot{v}}_{lv}}=\frac{{{{\dot{q}}}_{lv}}}{{{i}_{lv}}}\left( \frac{1}{{{\rho }_{v}}}-\frac{1}{{{\rho }_{l}}} \right)
\end{equation}
\begin{equation}\label{eq:eq13}
{{\dot{\alpha }}_{lv}}=-\frac{{{{\dot{q}}}_{lv}}}{\rho {{i}_{lv}}}
\end{equation}
The details of the phase change model can be found in Refs.\ \cite{Li2018BubbleEvaporation, Rattner2014VolumeOfFluidFormulation}.

\subsection{Simulation setup}\label{sec:sec023}
\begin{figure}
  \centering
  \includegraphics[scale=0.8]{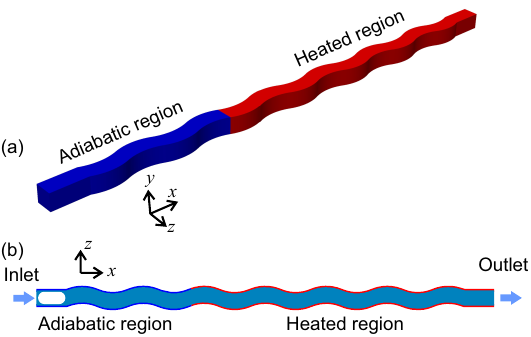}\\
  \caption{(a) 3D wavy microchannel; (b) Initial vapor bubble and the boundary conditions.}\label{fig:fig01}
\end{figure}

Fig.\ \ref{fig:fig01} shows the computational domain for simulating the flow boiling process in a wavy microchannel of a square cross-section with the side length of 1 mm. The channel length is 30 mm including 2-mm straight sections both at the inlet and the outlet. The wavy section spanned the middle 26-mm length of the channel with a hydraulic diameter, $D_h = 1$ mm. The shape of the two side walls follows the function of $z={\pm {{D}_{h}}}/{2}\;+A\sin [2\pi (x-{{x}_{0}})/\lambda ]$, where $\lambda $ and $A$ are the wavelength and the amplitude of the wavy channel, respectively, and $x_0= 2$ mm is the starting point of the wavy section. The channel is split into an adiabatic region of length $L_a = 10D_h$, used to allow the bubble to attain a steady-state motion, and a heated region of length $L_h = 20D_h$ where a uniform wall heat flux is applied. In this study, different waviness was considered by varying the amplitude ($A=$ 0, 0.1, 0.2, 0.3, 0.4 mm) and keeping the wavelength constant ($\lambda = 4$ mm). Therefore, the waviness of the channels $\gamma = A/\lambda $ varies from 0 to 0.1.

A vapor bubble of length $1.75D_h$ is initialized as a cylinder with spherical rounded caps at $0.1D_h$ from the channel inlet and an initial axial diameter of $0.8D_h$, as shown in Fig.\ \ref{fig:fig01}b. The initial volume of the bubble was fixed for all cases in order to establish a common baseline for comparison. The adiabatic region of the microchannel allows the bubble shape to deform and develop into an equilibrium shape before entering the heated region. A saturated liquid pushes the vapor bubble. The fluid is R134a with a saturation temperature of $T_{sat} = 30^\circ$C. The properties of the fluid are listed in Table \ref{tab:tab01} as obtained from Ref.\ \cite{Magnini2013ElongatedBubbleFlowBoiling, Potter2014R134a}.

\begin{table}[]
\centering
\small
\caption{Physical properties of the saturated fluids employed in this study.}\label{tab:tab01}
\begin{tabular}{|c|cc|cc|}
\hline
\multirow{2}{*}{Property}                        & \multicolumn{2}{c|}{R113}                                            & \multicolumn{2}{c|}{R134a}                                               \\ \cline{2-5}
                                                 & \multicolumn{1}{c|}{Liquid}                & Vapor                   & \multicolumn{1}{c|}{Vapor}                    & Vapor                    \\ \hline
Density,   $\rho$ {[}kg/m$^3${]}                 & \multicolumn{1}{c|}{1502}                  & 8                       & \multicolumn{1}{c|}{1187.5}                   & 37.535                   \\ \hline
Dynamic   viscosity, $\mu$ {[}Pa$\cdot$s{]}      & \multicolumn{1}{c|}{$4.77\times10^{-4}$}   & $1.04   \times 10^{-5}$ & \multicolumn{1}{c|}{$1.846   \times 10^{-4}$} & $1.238   \times 10^{-5}$ \\ \hline
Heat   capacity, $c_p$ {[}J/(kg$\cdot$K){]}      & \multicolumn{1}{c|}{943}                   & 695                     & \multicolumn{1}{c|}{1446}                     & 1065                     \\ \hline
Thermal   conductivity, $k$ {[}mW/(m$\cdot$K){]} & \multicolumn{1}{c|}{63.2}                  & 9.62                    & \multicolumn{1}{c|}{80.27}                    & 15.01                    \\ \hline
Saturation temperature, $T_{sat}$ {[}K{]}        & \multicolumn{2}{c|}{323.15}                                          & \multicolumn{2}{c|}{303.15}                                              \\ \hline
Surface tension, $\sigma$ {[}mN/m{]}             & \multicolumn{2}{c|}{14.4}                                            & \multicolumn{2}{c|}{7.56}                                                \\ \hline
Latent heat, $i_{lv}$ {[}kJ/kg{]}                & \multicolumn{2}{c|}{143.5}                                           & \multicolumn{2}{c|}{173.1}                                               \\ \hline
\end{tabular}
\end{table}

The velocity of the saturated liquid is imposed at the inlet, together with a zero-gradient condition for the pressure while the temperature is set equal to the saturation temperature $T = T_{sat}$. A zero gradient boundary condition is used at the outlet for both the velocity and the temperature, while the pressure is fixed at a reference value. The liquid mass flux at the inlet is $G=308$ kg/(m$^2\cdot$s), and a constant heat flux $q_w = 90$ kW/m$^2$ is applied at the wall of the heated region of the channel, which are typical values of mass flux and heat flux for thermal management. The constant-surface-heat-flux boundary is selected based on the assumption that the microchannel is heated by a source with known power per unit area, which is often a good approximation for flow boiling in microchannel applications, for example, power-dense electronics cooling.

Since the wavy microchannel has a larger channel length due to the waviness and a larger surface area for heat transfer, setting the same surface heat flux for the wavy microchannel will result in a larger total amount of heat transfer over the wavy microchannel than the straight channel. However, setting the same amount of heat transfer (i.e., reducing the heat flux for the wavy microchannel) is not ideal either, because for the same bubble size in microchannels with different waviness, the heat transfer to the region that the bubble covers will be different. Considering that the channel waviness does provide extra surface area for heat transfer in real applications, we set the same heat flux for different channels.

Dimensionless numbers are used to analyze the two-phase flow and heat transfer, including the Weber number ($\We$), the Capillary number ($\Ca$), and the Jakob number ($\Ja$). The Weber number indicates the ratio of the inertia to the surface tension forces, and it is defined as $\We = {{{D}_{h}}{{G}^{2}}}/({\rho \sigma })$; the Capillary number is the ratio of viscous to surface tension forces and it is defined as $\Ca = {\rho_l \nu_l U}/{\sigma }$; and the Jakob number is the ratio of the sensible heat to the latent heat of vaporization and it is defined as $\Ja = {q_w{{A}_{s}}}/({{{A}_{c}}U{{\rho }_{v}}i_{lv}})$. The parameters have been systematically varied while keeping all other properties constant. For the $\We$ number, the surface tension is varied from 0.04 to 0.0756 N/m. For the $\Ca$ number, the kinematic viscosity of the liquid is varied from $0.155\times10^{-6}$ to $4 \times10^{-6}$ m$^2$/s. For $\Ja$ number, the latent heat is varied from 140 to 200 kJ/kg.

The heat transfer on the wall of the microchannel is characterized by the Nusselt number, which is calculated from the heat transfer coefficient,
\begin{equation}\label{eq:eq14}
\Nu=\frac{h{{D}_{h}}}{{{k}_{l}}}
\end{equation}
\begin{equation}\label{eq:eq15}
h=\frac{q_w}{{{T}_{w}}-{{T}_{sat}}}
\end{equation}
where $T_w$ is the wall temperature extracted from the simulations, and $h$ is the heat transfer coefficient.

\subsection{Mesh independence analysis}\label{sec:sec024}
\begin{figure}
  \centering
  \includegraphics[scale=0.8]{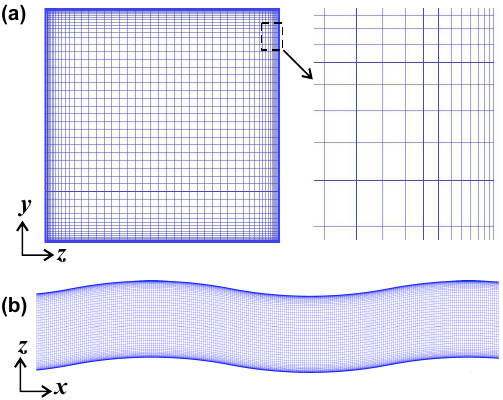}\\
  \caption{Mesh distribution for the simulation: (a) in the cross-section of the microchannel, (b) along the flow direction of the microchannel (only a section is shown).}
\label{fig:fig02}
\end{figure}

Prior to the numerical investigation on flow boiling heat transfer in wavy microchannels, a mesh independence study is performed to verify that the baseline mesh can sufficiently resolve the bubble growth in the microchannel and provides a moderate mesh scheme for balancing the numerical cost and accuracy. The domain is discretized with a structured mesh made of orthogonal hexahedral elements. In the cross-section, the core of the channel is uniformly meshed and a gradually refined mesh in the near-wall region, as shown in Fig.\ \ref{fig:fig02}(a). At least five cells are guaranteed for the near-wall region, as suggested by Gupta et al.\ \cite{Gupta2009ModellingTaylorFlow}. In the flow direction of the microchannel, the mesh distribution is uniform, as shown in Fig.\ \ref{fig:fig02}(b).

\begin{figure}
  \centering
  \includegraphics[scale=0.35]{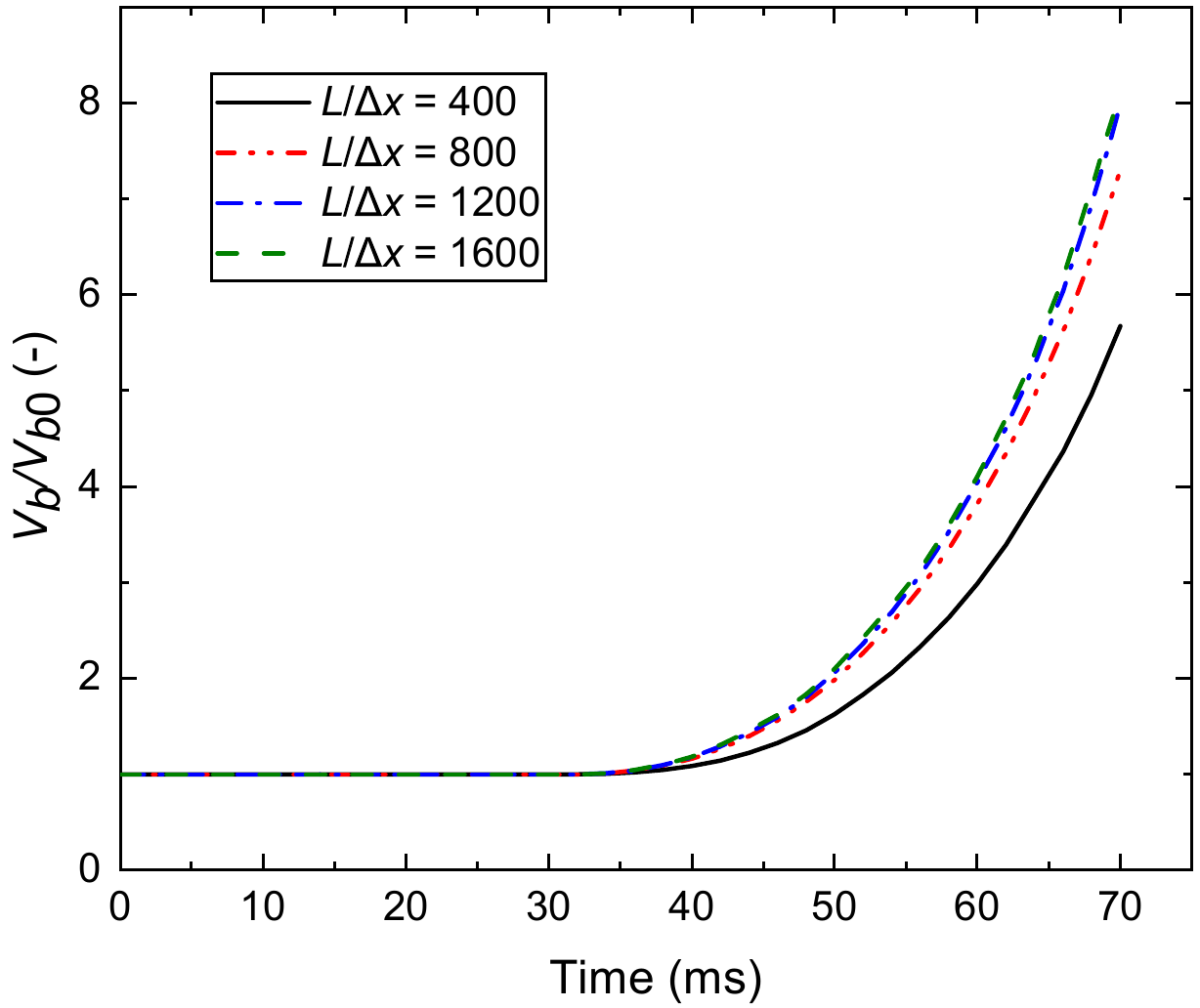}\\
  \caption{Non-dimensional bubble volume (i.e., ${{V}_{b}}/{{V}_{b0}}$, where ${{V}_{b}}$ is the bubble volume and ${{V}_{b0}}$ is the initial bubble volume) plotted as a function of time under various mesh sizes.}
\label{fig:fig03}
\end{figure}

The mesh independence study spanned four mesh resolutions. Numerical cases are conducted for a wavy channel of $\gamma = 0.025$ with various mesh resolutions of $L/\Delta x =$ 400, 800, 1200, and 1600, where $L$ is the length of the microchannel and $\Delta x$ represents the mesh size in $x$ direction. Fig.\ \ref{fig:fig03} shows the results of the mesh-independence test in terms of non-dimensional bubble volume plotted as a function of time under various mesh sizes. Evaporation starts after the bubble mass center reaches the heated region at time $t= 35$ ms. The results of $L/\Delta x = 400$ and 800 show a large deviation from the results of $L/\Delta x = 1200$ and 1600. Although there are still minute deviations among the results of $L/\Delta x = 1200$ and 1600 due to different mesh resolutions, the mesh with $L/\Delta x = 1200$ is able to capture the bubble growth accurately. Therefore, it is chosen for the subsequent simulations.

\subsection{Validation}\label{sec:sec025}
\begin{figure}
  \centering
  \includegraphics[scale=0.35]{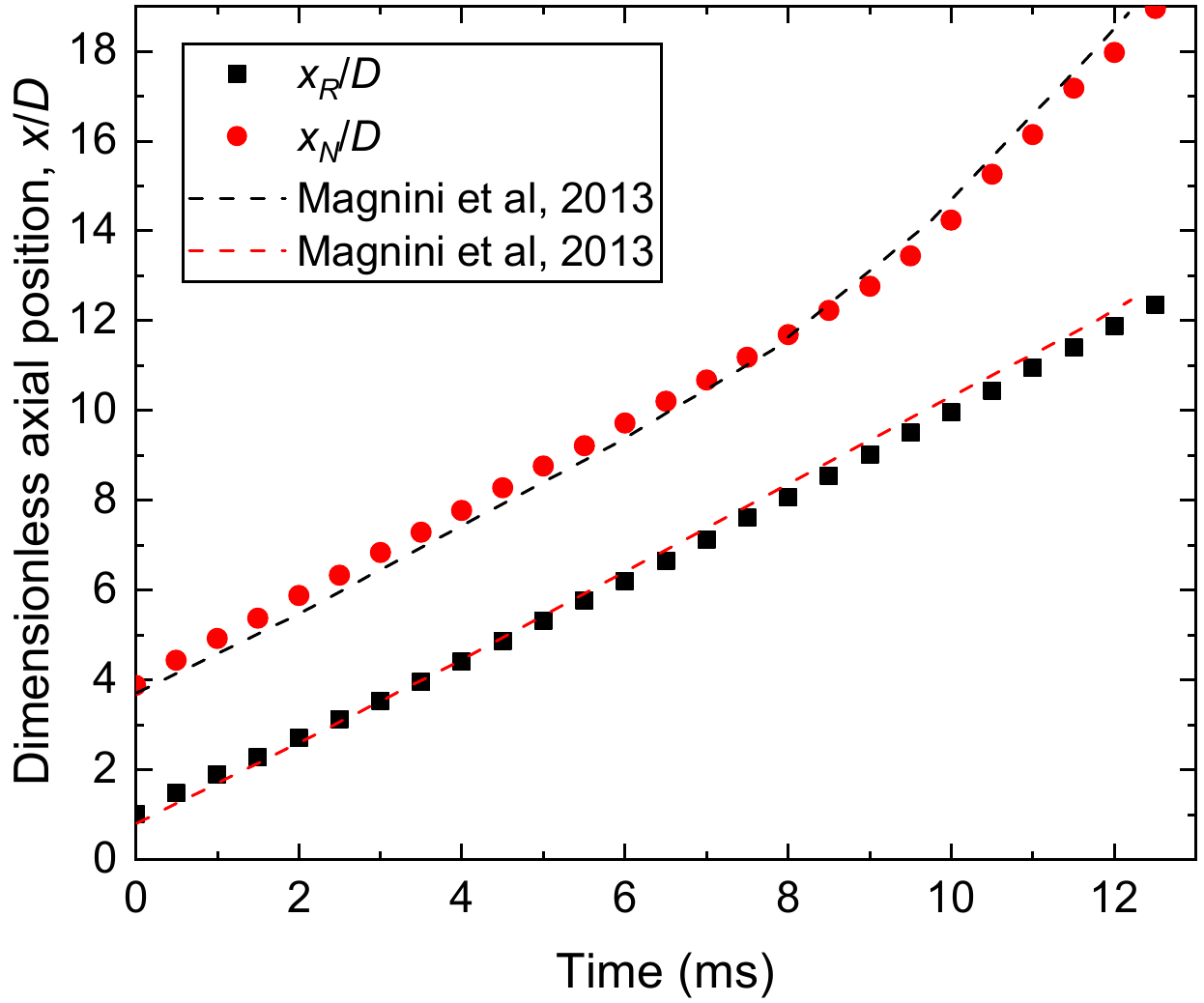}\\
  \caption{Dimensionless bubble positions (rear $x_R$ and nose $x_N$) as a function of time compared with the results obtained by Magnini et al.\ \cite{Magnini2013ElongatedBubbleFlowBoiling}.}
\label{fig:fig04}
\end{figure}

In order to verify the accuracy of the phase change model, we performed a simulation of an elongated bubble under flow boiling conditions in a circular microchannel and the results are compared with those obtained by Magnini et al.\ \cite{Magnini2013ElongatedBubbleFlowBoiling} under the same flow conditions. The channel is modeled as a two-dimensional axisymmetric computational domain, which is a simplification of an elongated bubble placed at the center of a circular microchannel with diameter $D = 0.5$ mm and length $L = 20D$, with no gravity effect. The channel is split into an adiabatic region of length $L_a = 8D$ and a heated region of length $L_h = 12D$ where a constant wall heat flux, $q_w = 9$ kW/m$^2$, is imposed at the channel wall. Refrigerant R113 is used as the working fluid, which enters the microchannel with a uniform inlet velocity of 0.4 m/s and a temperature of 323.15 K. An elongated vapor bubble of length $3D$ is initialized in the adiabatic section as a cylinder with spherical rounded caps on both sides at $1D$ from the channel inlet. The fluid properties are summarized in Table \ref{tab:tab01}. The bubble positions (rear $x_R$ and nose $x_N$) are compared with the results obtained by Magnini et al.\ \cite{Magnini2013ElongatedBubbleFlowBoiling} in Fig.\ \ref{fig:fig04}, which shows good agreement.
The maximum discrepancy with the simulations is 7.5\% for rear $x_R$ and 5.2\% for nose $x_N$. The difference may be due to the numerical algorithm or the discretization scheme used in the simulation. But overall, it is small for the current problem, indicating a good accuracy of the simulation.

\section{Results and discussion}\label{sec:sec03}
\subsection{Effect of waviness of the microchannel}\label{sec:sec031}
\begin{figure}
  \centering
  \includegraphics[scale=0.6]{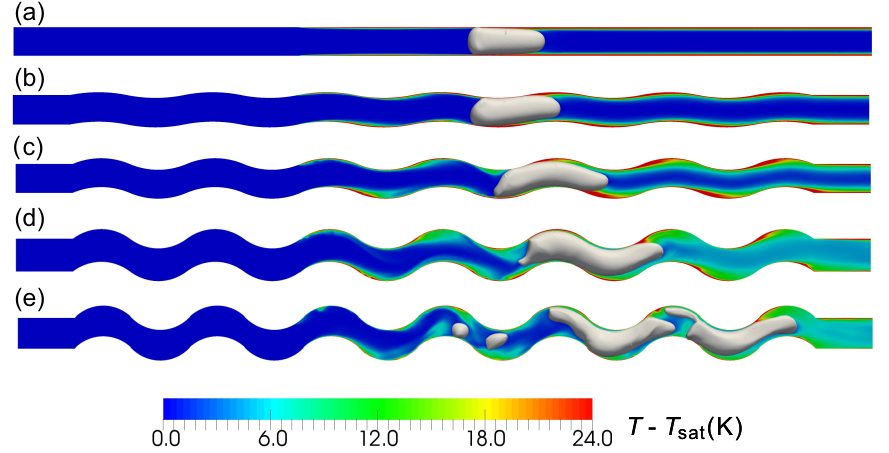}\\
  \caption{Bubble shapes and temperature field $t = 52$ ms for microchannels with different waviness taken at the middle cross-section in $y$-direction. (a) $\gamma= 0$, (b) $\gamma=  0.025$, (c) $\gamma=  0.05$, (d) $\gamma=  0.075$, and (e) $\gamma=  0.1$.}
\label{fig:fig05}
\end{figure}

At the inception of the flow, the saturated liquid enters the microchannel and transports the bubble downstream. Evaporation starts when the bubble interface gets in contact with the superheated thermal boundary layer developing at the heated wall. As liquid turns into vapor, the bubble nose accelerates downstream and elongates. The growth and motion of bubbles in both straight and wavy microchannels are compared in this section. Fig.\ \ref{fig:fig05} shows the growth of a vapor bubble and the temperature field inside the straight and wavy microchannels at a typical instant. It can be seen that as the waviness becomes larger, the bubbles become larger with stronger deformation. For microchannels with very large waviness, the bubble may break up as it is transferred to downstream, as shown in Fig.\ \ref{fig:fig05}(e). The breakup of the bubble could be attributed to the bubble deformation induced by the wavy channel and the flow velocity in the perpendicular direction, which could facilitate the development of the surface instability and the pinch-off of the bubble.

\begin{figure}
  \centering
  \includegraphics[scale=0.6]{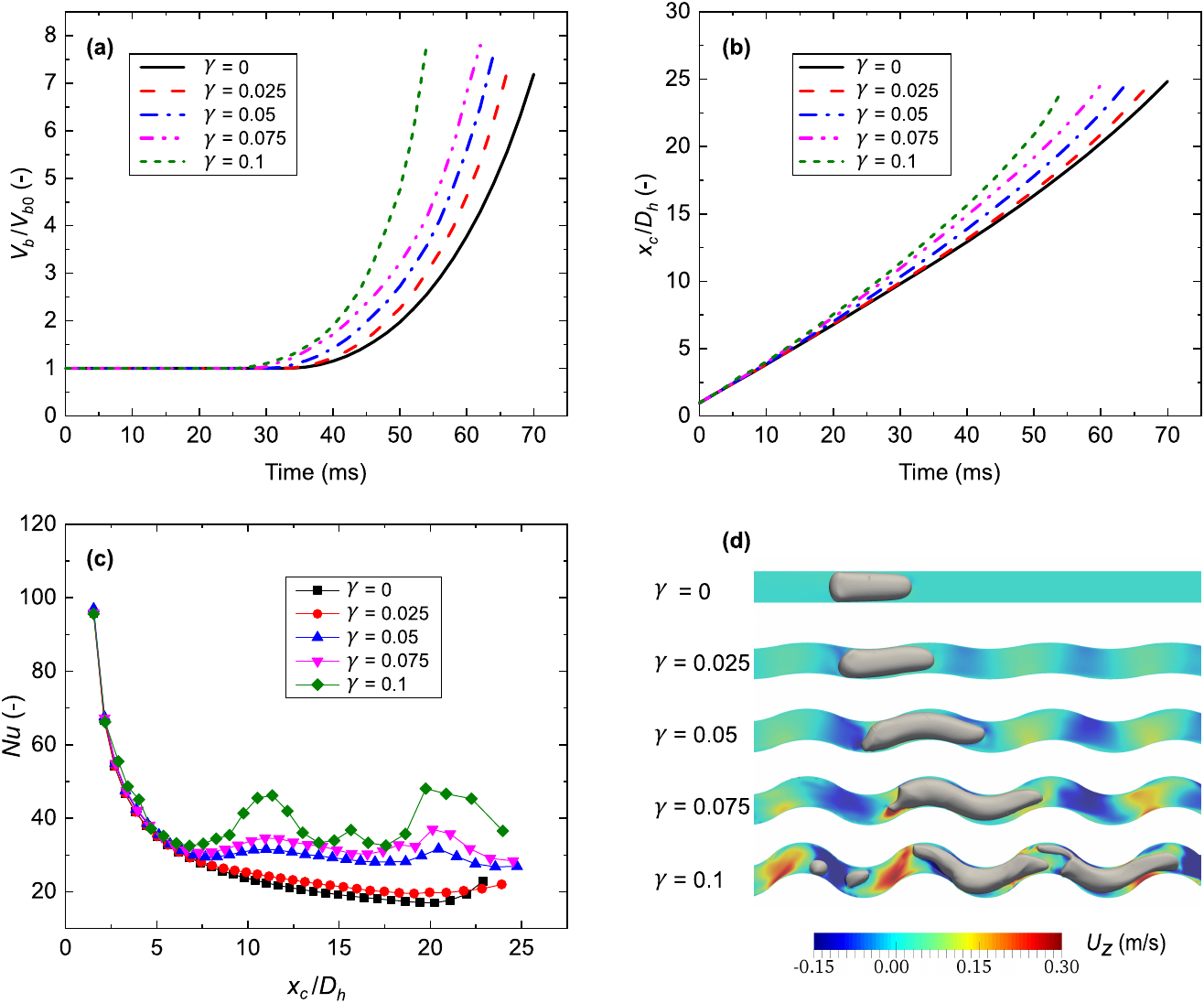}
  \caption{(a) Dimensionless bubble volume $V_b/V_{b0}$, plotted as a function of time for different waviness ($\gamma$). (b) Dimensionless bubble position $x_c/D_h$ as a function of time for different waviness. (c) Nusselt number at the bottom wall, extracted at mid-point of $x = 21D_h$ against dimensionless bubble position $x_c/D_h$. (d) Distribution of velocity component in $z$-direction $u_z$ near the bubbles at $t = 52$ ms for microchannels with different waviness taken at the middle cross-section in $y$-direction.
  }
\label{fig:fig06}
\end{figure}

To further quantify the effect of the waviness on the bubble motion and the corresponding wall heat transfer, the dimensionless bubble volume, the bubble position, and the Nusselt number are plotted in Figs.\ \ref{fig:fig06}(a), \ref{fig:fig06}(b), and \ref{fig:fig06}(c), respectively. Even though a constant surface heat flux is specified on the wall of the microchannel, the growth rate of the bubble size is not constant. When the bubble is in the adiabatic region, the bubble size does not change, and when the bubble reaches the heated region, the bubble grows in size gradually. The growth rate of the bubble size increases as the bubble moves to downstream. This is because as the bubble size increases, the interaction area between the bubble and the wall of the microchannel becomes larger. Therefore, the heat transfer from the wall to the bubble is enhanced, speeding up the growth rate of the bubble size. Fig.\ \ref{fig:fig06}(a) also shows that the bubble growth rate is larger for microchannels with larger waviness. This could be attributed to the extra perturbation to the flow induced by the waviness of the microchannel. The waviness of the microchannel can induce flow velocity perpendicular to the direction of the microchannel as shown in Fig.\ \ref{fig:fig06}(d). As the waviness increases, the velocity perpendicular to the direction of the microchannel becomes larger. Such velocity component can transfer high-temperature fluid from the wall to the central region, and enhance the heat transfer from the wall \cite{Che2012HeatTransfer2DTA}. In addition, the waviness of the microchannel affects the thickness of the liquid film between the bubble and the wall. The bubble tends to maintain a straight shape by the surface tension. Thus, the liquid film becomes thinner near the convex region of the wall. Because the heat transfer from the wall to the bubble is mainly by the conduction through the liquid film, the local thin film reduces the local thermal resistance and can enhance the heat transfer from the wall to the bubble. As a consequence, the bubble grows faster for microchannels with larger waviness.

The position of the bubble in the microchannel is plotted in Fig.\ \ref{fig:fig06}(b). Initially, the bubble moves forward at a constant speed. After the bubble reaches the heated region, the bubble accelerates. And the acceleration is affected by the waviness of the microchannel. For microchannels with higher waviness, the acceleration is higher. The main reason for the bubble acceleration is the growth of the bubble size as the bubble moves forwards. For microchannels with higher waviness, the bubble is larger due to the enhanced heat transfer, and the bubble position is accelerated by the bubble expansion.

The Nusselt number at a typical point of the microchannel is plotted versus the dimensionless bubble position $x_c/D_h$  in Fig.\ \ref{fig:fig06}(c). The local Nusselt number at a point is selected to show the transient effect of the passing-by of the bubble, which can be used to explain the mechanism of heat transfer enhancement. The Nusselt number is very large initially and quickly decreases. This could be attributed to the large initial temperature gradient at the wall of the microchannel. As the initial temperature gradient disappears, the Nusselt number quickly decreases. Then the Nusselt number for the straight microchannel gradually decreases, even the passing through of the bubble does not produce noticeable perturbation to the Nusselt number. This is mainly because of the thick liquid film between the bubble and the wall as the bubble is small and the wall is flat. In contrast, as the waviness increases, the bubble becomes larger and the wall becomes wavier. Therefore, the heat transfer is significantly enhanced and is perturbed by the passing through of the vapor bubble. As a consequence, the Nusselt number for microchannels with larger waviness is much larger and, as the bubble passes through, exhibits stronger fluctuation.
When the waviness is very large (e.g., $\gamma = 0.1$), the large vapor bubble may break up into several pieces (as shown in Fig.\ \ref{fig:fig05}(e)). The sequential pass-through of the small bubbles of different sizes will lead to strong fluctuation of the Nusselt number (as shown in Fig.\ \ref{fig:fig06}(c)), hence effectively enhancing the heat transfer. The Nusselt number achieved for $\gamma = 0.1$ is peaking at 48.1 compared to 18.8 for the straight microchannel, i.e., 2.6 times higher. These results of the Nusselt number also confirm that the heat transfer enhancement in the wavy microchannel is not due to the larger surface area for heat transfer, but due to the flow perturbation when the bubble passes through the wavy channel.

\subsection{Effect of Weber number}\label{sec:sec032}
\begin{figure}
  \centering
  \includegraphics[scale=0.6]{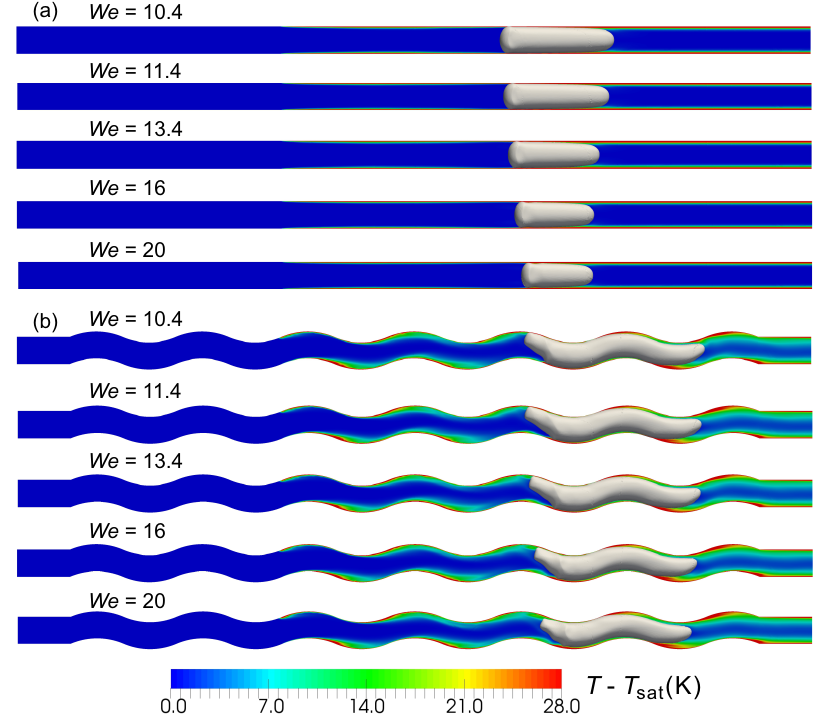}\\
  \caption{Temperature field for different $\We$ numbers at 60 ms taken at the middle cross-section in $y$-direction: (a) for straight microchannel $\gamma= $ 0, (b) for wavy microchannel $\gamma= $ 0.05. }
\label{fig:fig07}
\end{figure}

\begin{figure}
  \centering
  \includegraphics[scale=0.55]{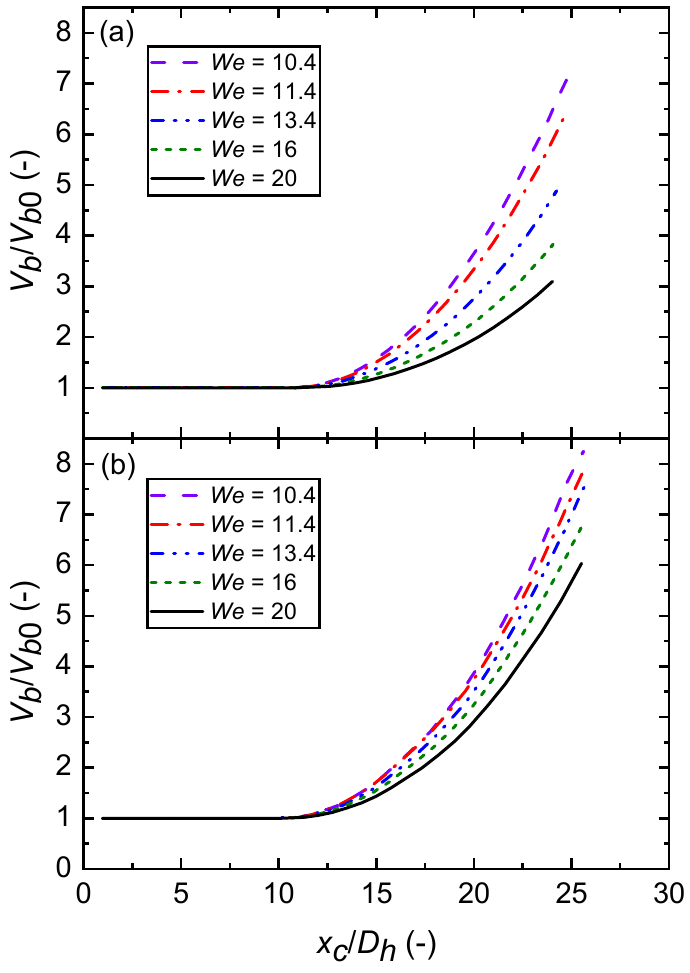}\\
  \caption{Dimensionless bubble volume, $V_b/V_{b0}$, as a function of the dimensionless bubble mass center position, $x_c/D_h$, for different values of Weber number. (a) $\gamma=  0$ and (b) $\gamma=  0.05$.}
\label{fig:fig08}
\end{figure}

The Weber number is an important parameter to study the relative effects of surface tension and inertial forces on flow patterns in microchannels \cite{Kandlikar2004MicrochannelFlowBoiling}. In this study, the $\We$ number was varied by changing the surface tension. The temperature field with the bubble flow pattern is presented in Fig.\ \ref{fig:fig07} for different $\We$ numbers, and the corresponding bubble volumes are quantitatively compared in Fig.\ \ref{fig:fig08}. It can be seen that the Weber number has a significant impact on the bubble growth and the effect is different between straight and wavy microchannels.

For the straight microchannel, the bubble becomes smaller as the Weber number increases. The Weber number has a direct impact on the bubble shape as it moves to downstream. As the Weber number increases, the relative importance of the surface tension effect reduces. Since the surface tension tends to restore the bubble to a spherical shape, the bubble with larger surface tension has a thinner liquid film between the bubble and the wall. In contrast, the bubble with smaller surface tension is highly deformed and has a thicker liquid film between the bubble and the wall. Therefore, at a larger Weber number, the thermal resistance between the wall and the bubble is larger, and the heat transfer from the wall to the bubble is reduced as the heat transfer is mainly by thermal conduction. As a consequence, the growth rate of the bubble at a higher Weber number is smaller.

The effect of the Weber number for the wavy microchannel shares similar features to that for the straight channel: the bubble growth rate is smaller for larger Weber numbers, but the effect of the Weber number is weaker than that in the straight microchannel, as shown in Fig.\ \ref{fig:fig08}. This is because the waviness affects the film thickness. The wavy wall of the microchannel deforms the shape of the liquid film, and increases the heat transfer from the wall to the bubble significantly. Therefore, the relative effect of the Weber number becomes weaker than the straight microchannel.

\subsection{Effect of Capillary number}\label{sec:sec033}
\begin{figure}
  \centering
  \includegraphics[scale=0.6]{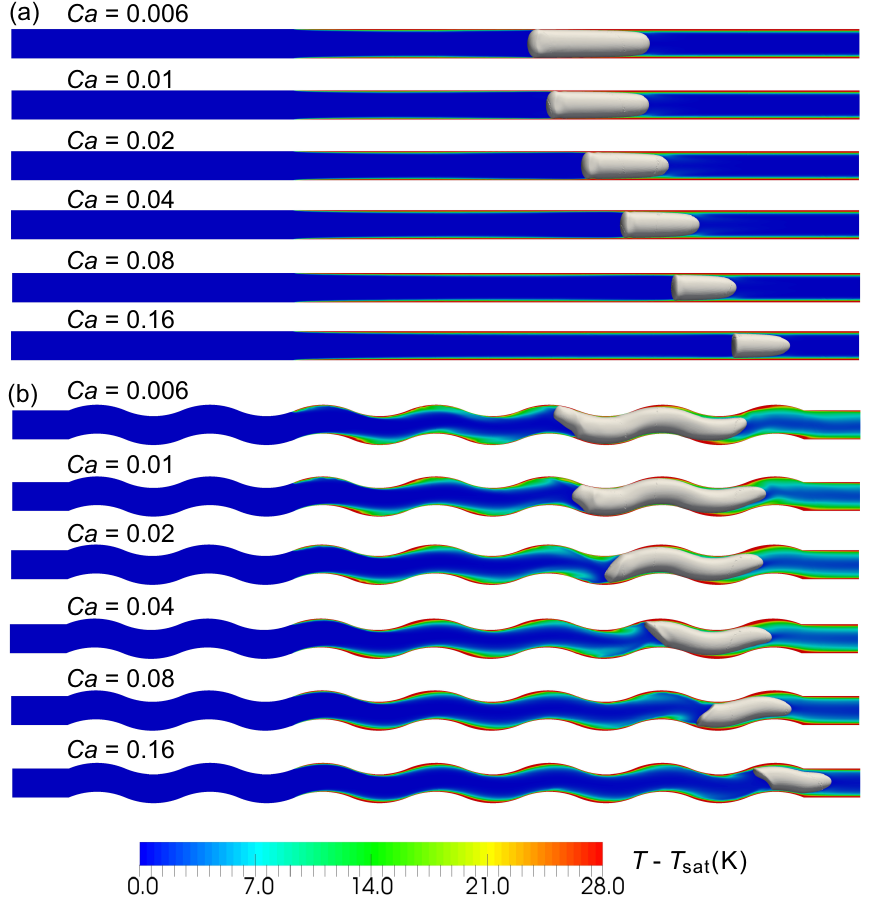}\\
  \caption{Temperature field for different $\Ca$ numbers at 60 ms taken at the middle cross-section in $y$-direction: (a) for straight microchannel $\gamma= 0$, (b) for wavy microchannel $\gamma=  0.05$.}
\label{fig:fig09}
\end{figure}

\begin{figure}
  \centering
  \includegraphics[scale=0.5]{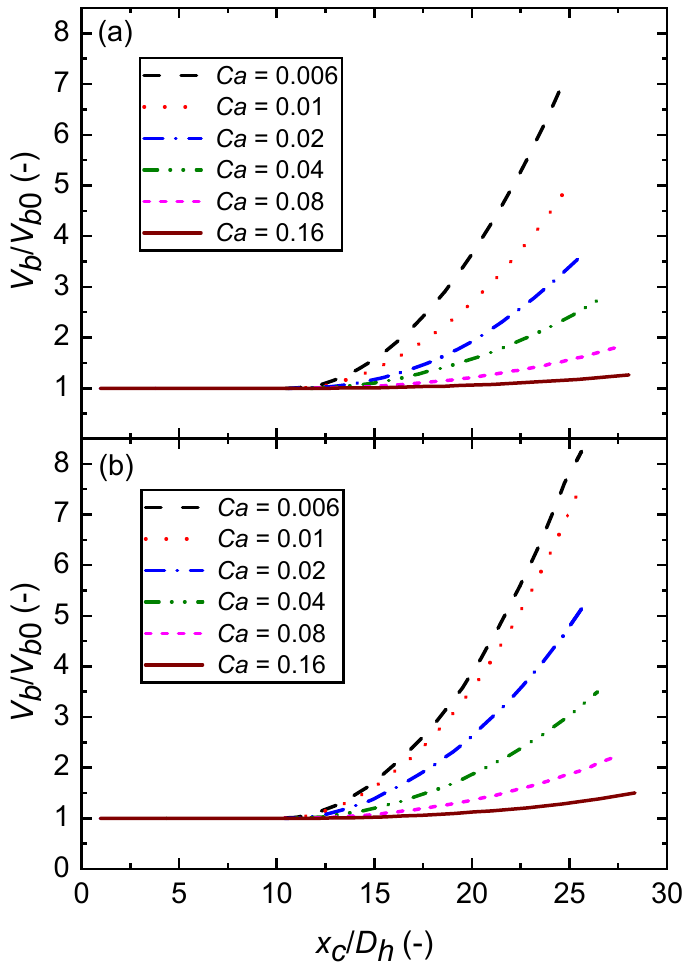}\\
  \caption{Dimensionless bubble volume, $V_{b}/V_{b0}$, as a function of the dimensionless bubble mass center position, $x_c/D_h$, for different Capillary numbers. (a) $\gamma=  0 $ and (b) $\gamma= 0.05. $ }
\label{fig:fig10}
\end{figure}

The Capillary number represents the ratio of viscous to surface tension forces and significantly affects the bubble dynamics \cite{Kandlikar2004MicrochannelFlowBoiling}. Here, the effect of the $\Ca$ number on the boiling heat transfer in microchannels is studied by changing the viscosity of the liquid. The temperature field with bubble flow pattern is shown in Fig.\ \ref{fig:fig09} for different $\Ca$ numbers, and the corresponding bubble volumes are quantitatively compared in Fig.\ \ref{fig:fig10}. The effects of the $\Ca$ number on bubble growth inside the straight and wavy microchannels are significant. By increasing the $\Ca$ number, the growth rate of the bubble size decreases remarkably for both straight and wavy microchannels. The more viscous the fluid is, the slower the bubble grows. This is because as the $\Ca$ number increases, the viscous force becomes larger, and the relative effect of the surface tension decreases. Since the surface tension has the effect of restoring the bubble into the spherical shape, the reduction in the surface tension effect leads to the thickening of the liquid film between the bubble and the wall. As a consequence, the heat transfer from the wall to the bubble is suppressed by the thick liquid film. Therefore, the growth rate of the bubble size is smaller for higher Capillary numbers.

\begin{figure}
  \centering
  \includegraphics[scale=0.5]{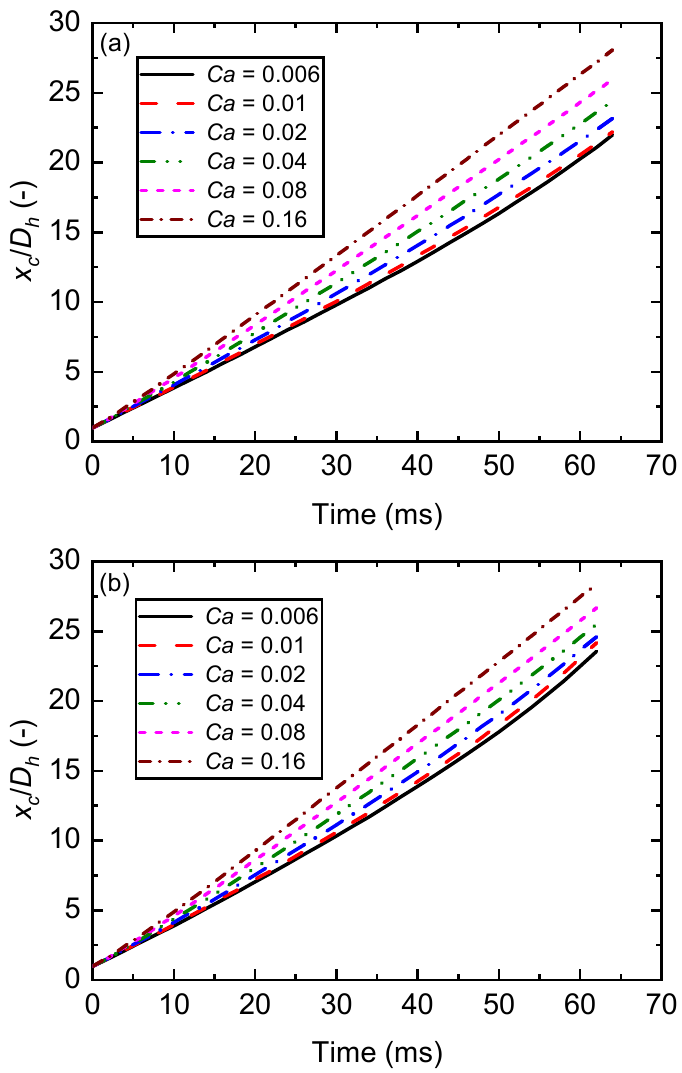}\\
  \caption{Dimensionless bubble position $x_c/D_h$ as a function time for different Capillary numbers. (a) $\gamma= 0$  and (b) $\gamma=  0.05 $.}
\label{fig:fig11}
\end{figure}

The bubble position is also significantly affected by the Capillary number, as shown qualitatively in Fig.\ \ref{fig:fig09} and also shown quantitatively in Fig.\ \ref{fig:fig11}. For bubbles in microchannels with smaller $\Ca$, the bubbles move slower, and the bubble speed increases as $\Ca$ increases. This is mainly due to the effect of the film thickness on the bubble speed. For flow with larger $\Ca$, the bubble size is much smaller, and the film between the bubble and the wall is much thicker. As the flow is laminar, the flow speed in the center of the microchannel is much larger than that near the wall. Therefore, the speed of the bubble is higher than the average speed of flow in the microchannel, and the smaller the bubble is, the higher the bubble speed is. Therefore, smaller bubbles move faster in the microchannel. Even though the bubbles at higher $\Ca$ are smaller and expand slower than that at small $\Ca$, the effect of the film thickness on bubble speed is significant, resulting in faster bubble speeds than that at small $\Ca$.

\subsection{Effect of Jakob number}\label{sec:sec034}
\begin{figure}
  \centering
  \includegraphics[scale=0.6]{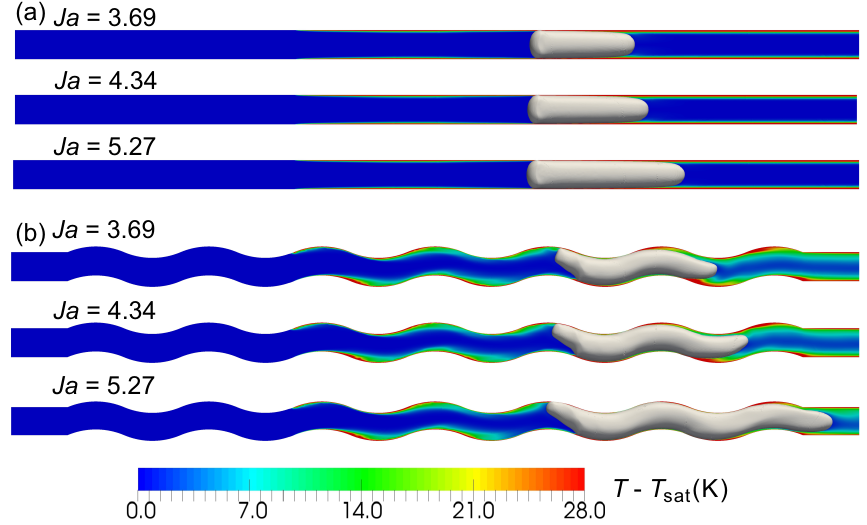}\\
  \caption{Temperature field for different $\Ja$ numbers at 60 ms taken at the middle cross-section in $y$-direction: (a) for straight microchannel $\gamma=  0$, (b) for wavy microchannel $\gamma=  0.05. $}
\label{fig:fig12}
\end{figure}

\begin{figure}
  \centering
  \includegraphics[scale=0.5]{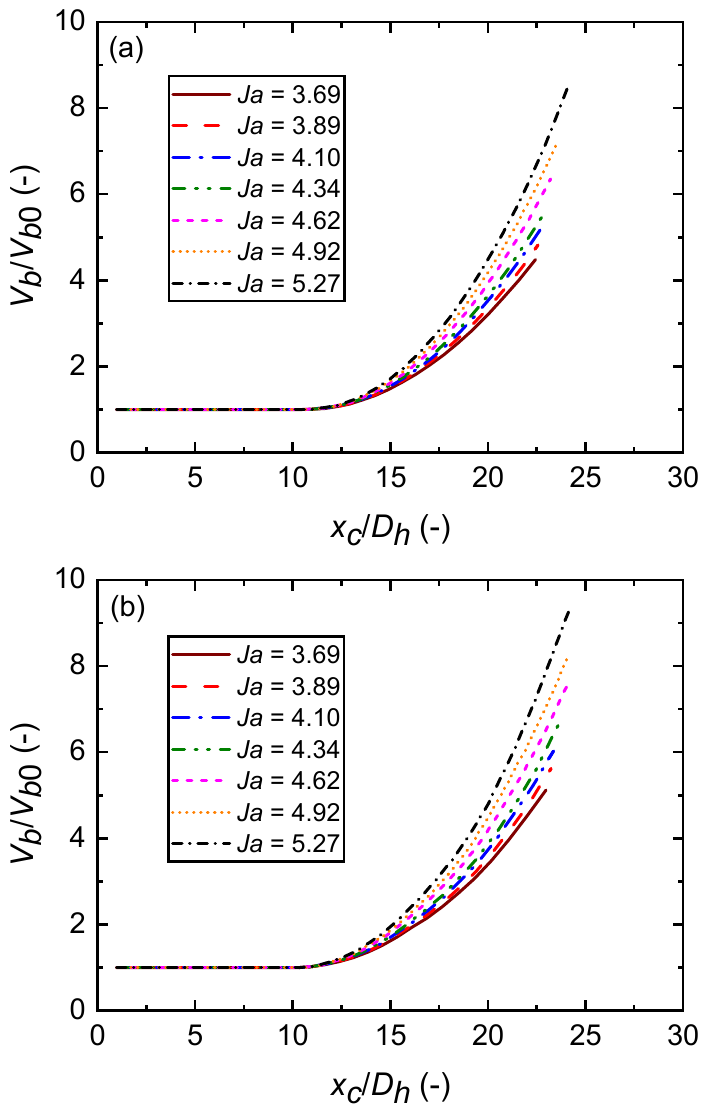}\\
  \caption{Dimensionless bubble volume, $V_b/V_{b0}$, as a function of the dimensionless bubble mass center position, $x_c/D_h$, for different values of Jakob number. (a) $\gamma=  0$ and (b) $\gamma=  0.05$.}
\label{fig:fig13}
\end{figure}

The Jakob number represents the ratio of the sensible heat to the latent heat of vaporization. Here, it is varied by changing the latent heat of the fluid. The temperature field with bubble flow pattern is presented in Fig.\ \ref{fig:fig12} for different $\Ja$ numbers at 60 ms. The effects of Jakob number on the bubble size in straight and wavy microchannels are shown in Fig.\ \ref{fig:fig13}. By increasing the $\Ja$ number, the bubble volume increases moderately for both straight and wavy microchannels. As the $\Ja$ number increases, the latent heat of vaporization decreases. Therefore, with the same amount of heat absorbed from the wall, more fluid will vaporize, and produces larger bubbles. With the presence of waviness, the effect of the $\Ja$ number is even stronger as compared with straight microchannels, as shown in Fig.\ \ref{fig:fig13}(b).

\section{Conclusions}\label{sec:sec04}
In this study, we numerically investigate the flow and heat transfer of vapor bubbles during flow boiling inside wavy microchannels. The numerical model is validated against previous results obtained for straight microchannels. The effects of important dimensionless numbers are investigated, including the Weber number, the Capillary number, and the Jakob number. The results show that the waviness of the microchannel can significantly affect the dynamics of the vapor bubble and the heat transfer performance. As the waviness increases, the bubble grows faster because of the bubble deformation by the wavy channel. In addition, the bubble moves faster due to the bubble expansion, and heat transfer is enhanced by the perturbation of the bubble. Furthermore, the results provide a novel insight into the influence of velocity perpendicular to the flow direction of the microchannel as the waviness increases, which transfers high-temperature fluid from the wall to the central region. The local Nusselt number of the wavy channel could be high up to 2.6 times higher than the straight channel. As the Weber number increases, the bubble becomes smaller because of the weaker effect of the surface tension and the thicker liquid film between the bubble and the wall. The relative effect of the Weber number is weaker for wavy microchannels than for straight microchannels. By increasing the Capillary number, the growth rate of the bubble size decreases remarkably because the smaller surface tension leads to the thickening of the liquid film between the bubble and the wall and suppresses the heat transfer. In addition, the bubble speed becomes faster because of the smaller bubble size and the thicker liquid film. By increasing the Jakob number, the bubble volume increases for both straight and wavy microchannels because of the decrease in the latent heat of vaporization.

This study focuses on the bubble growth and the heat transfer enhancement by considering only a single channel. Real microchannel heat sinks usually consist of many microchannels in parallel. More work is required in this field to consider the interaction between channels, or the heat transfer process in the entire heat sink, which is computationally expensive. The current study not only provides insight into the mechanism of heat transfer enhancement, but also is helpful for the design of microchannel sinks.

\section*{Declaration of Competing Interest}
None.

\section*{Acknowledgements}
This work was supported by the Key Laboratory Fund (6142702200509) and the National Natural Science Foundation of China (Grant Nos.\ 52176083 and 51921004).



\bibliography{BoilingHeatTransfer}

\end{document}